\documentclass[%
reprint,
superscriptaddress,
frontmatterverbose, 
 amsmath,amssymb,
 aps,
]{revtex4-1}
\usepackage{xcolor}
\usepackage{graphicx}
\usepackage{dcolumn}
\usepackage{bm}
\usepackage{footmisc}

\usepackage{mathtools}
\usepackage{esvect}

\DeclareMathOperator\erf{erf}
\begin{document}

\preprint{APS/123-QED}

\title{Frequency Shifts due to Stark Effects on a Rb two-photon transition}
\thanks{A contribution of AFRL, an agency of the US government, and not subject to copyright in the United States.}%

\author{Kyle W. Martin}
 \affiliation{Applied Technology Associates dba ATA,
 1300 Britt Street SE, Albuquerque, NM 87123}
 
   \author{Benjamin Stuhl}
\affiliation{%
 Space Dynamics Laboratory,
 1695 North Research Park Way,  
 North Logan, Utah 84341
}%

   \author{Jon Eugenio}
\affiliation{%
Stevens Institute of Technology,
1 Castle Point Terrace, 
Hoboken, NJ 07030
}%

\author{Marianna S. Safronova}
\affiliation{%
University of Delaware
217 Sharp Lab,
Newark, DE 19716
}%
\affiliation{Joint Quantum Institute, National Institute of Standards and Technology and the University of Maryland, College Park, Maryland 20742, USA}

\author{Gretchen Phelps}%
\affiliation{%
 Air Force Research Laboratory, 
 Space Vehicles Directorate, 
 Kirtland Air Force Base, NM 87117
}%

\author{John H. Burke}
\affiliation{%
 Air Force Research Laboratory, 
 Space Vehicles Directorate,
 Kirtland Air Force Base, NM 87117
}%

\author{Nathan D. Lemke}
\affiliation{%
Bethel University
3900 Bethel Drive, 
St. Paul, MN 55112
}%

\date{\today}

\begin{abstract}

\noindent The $5S_{1/2}\rightarrow 5D_{5/2}$ two-photon transition in Rb is of interest for the development of a compact optical atomic clock. Here we present a rigorous calculation of the 778.1~nm ac-Stark shift ($2.30(4) \times10^{-13}$(mW/mm$^2$)$^{-1}$) that is in good agreement with our measured value of $2.5(2) \times10^{-13}$(mW/mm$^2$)$^{-1}$.  We include a calculation of the temperature-dependent blackbody radiation shift, we predict that the clock could be operated either with zero net BBR shift ($T=495.9(27)$~K) or with zero first-order sensitivity ($T=368.1(14)$~K).   Also described is the calculation of the dc-Stark shift of 5.5(1)$\times 10^{-15}$/(V/cm$^2$)  as well as clock sensitivities to optical alignment variations in both a cat's eye and flat mirror retro-reflector.  Finally, we characterize these Stark effects discussing mitigation techniques necessary to reduce final clock instabilities.
\end{abstract}

\maketitle


\section{Introduction}

Narrow-line transitions can be realized through two-photon spectroscopy to explore a wide array of scientific phenomena.   Two-photon transitions have been successfully leveraged for measuring fine  and hyperfine structures 
\cite{HANSCH1974}, Zeeman 
\cite{Salour1974} and Stark 
\cite{Schawlow1975} splittings, hot vapour collisional effects 
\cite{Grynberg1975,Zameroski2014}, and are important in precision Hydrogen spectroscopy 
\cite{Parthey2011}.  A common technique in two-photon spectroscopy is the degenerate two-photon method which uses photons derived from the same source, resulting in identical frequencies.
The advantage of using degenerate photons in two beams of opposite $\vv{\bm{k}}$ vector is that all atoms, regardless of velocity, can contribute to a Doppler-free signal \cite{Biraben1973,Grynberg1974,Bloembergen1974,Liao1974,HANSCH1974}. As employed here 
\cite{Martin2018} 
the Rb $5S_{1/2}\rightarrow 5D_{5/2}$ transition can be leveraged to create a high stability optical clock with a simple vapour cell architecture, without
the need for laser cooling, offering an alternative to saturated absorption systems such as molecular iodine 
\cite{Schkolnik2017}, or pulsed optically pumped microwave systems \cite{Emeric2017,Levi2012,Lin2017}.  
Offering a simpler and more compact approach than more complicated (albeit higher stability) optical lattice clocks  
\cite{Mullin1989,Brusch2006,Dzuba2010,Guo2010,Derevianko2011,Ludlow2015};  Doppler free degenerate two-photon spectroscopy provides an appealing architecture for which to build a compact optical atomic frequency standard 
\cite{Martin2018, Perrella2013, Phelps2018} and has already shown promise for very small packaging \cite{Newman:19}.

\begin{figure}
\includegraphics[width=0.47\textwidth]{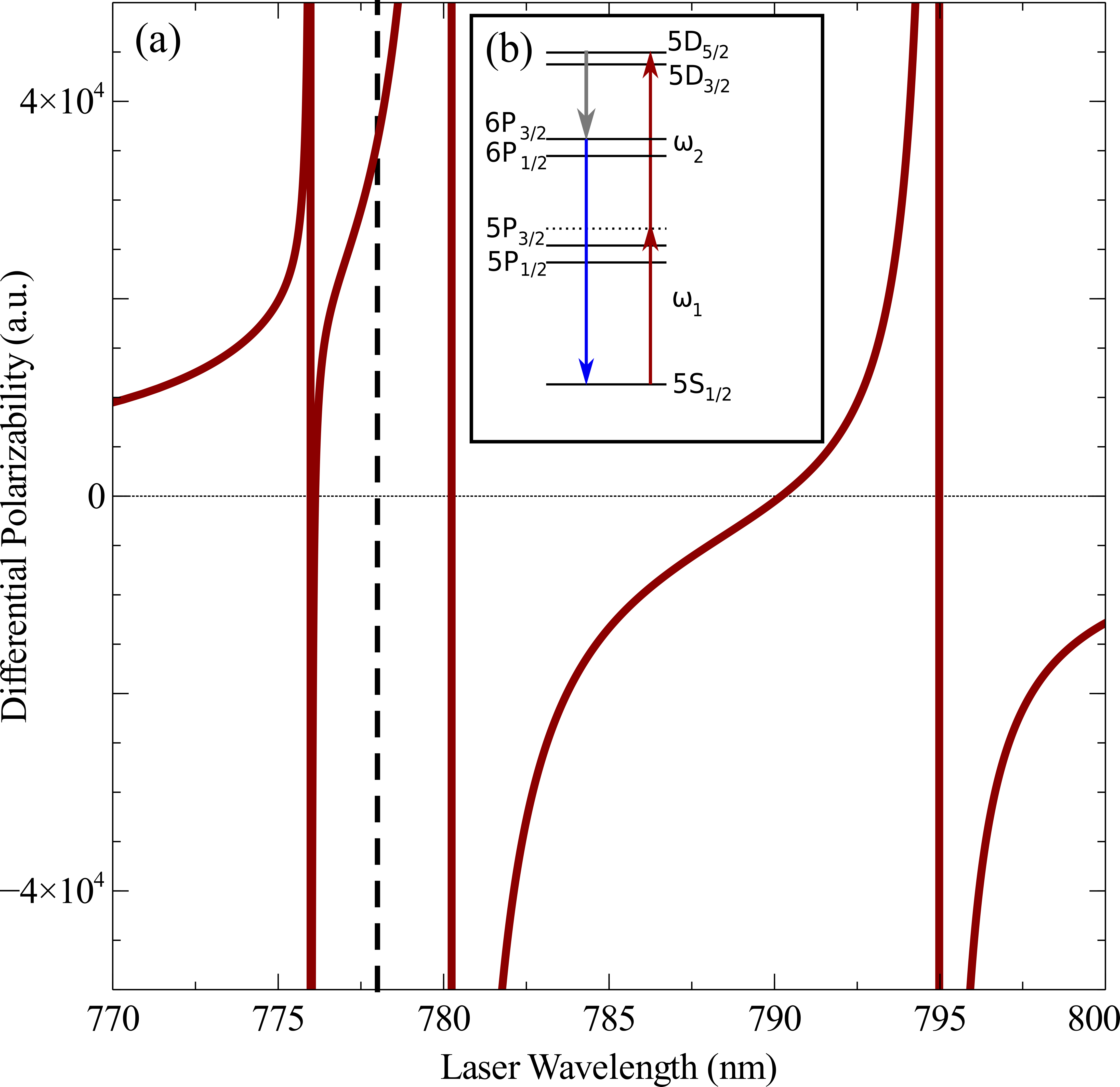}
\caption{\label{fig:twophoton} (a) The differential polarizability displayed in atomic units of the $5S_{1/2}\rightarrow 5D_{5/2}$ transition as a function of wavelength as calculated by Equation \ref{eq:scalepol}. (b) Partial energy level diagram of Rb. The virtual state, through which the two-photon transition is excited, is denoted by a horizontal dashed line.  In the degenerate case, $\omega_1=\omega_2$, the virtual state is 2~nm detuned from the $5P_{3/2}$ state.   The cascade decay path $5D_{5/2}\rightarrow$ $6P_{3/2} \rightarrow$ $5S_{1/2}$ results in the emission of a 420~nm photon, which we detect to observe the two-photon resonance. } 
\end{figure}




The two-photon transition in Rb can be driven with two degenerate 778.1~nm photons, which can be generated either directly with a 778.1~nm laser diode or through second harmonic generation (SHG) of mature telecommunications C-band lasers at 1556.2~nm 
\cite{Martin2018}.  The relatively small virtual state detuning (see Figure \ref{fig:twophoton}) results in significant atomic excitation rates at modest optical intensities 
\cite{Nez1993,Touahri1997,Hilico1998}, which allows for a high signal-to-noise ratio of the 420~nm fluorescence, as shown in Figure \ref{fig:twophoton}, due to the simple detection scheme with spectral filtering of the incident probe beam from the fluorescence signal.  Fractional frequency instabilities of  $ < 4\times 10^{-13}/\sqrt{\tau}$ up to 10,000 seconds have been measured, and operation with increased SNR has shown short term stability of $1\times 10^{-13}$ at one second \cite{Martin2018}.  
This performance is comparable to other commercially available compact clocks, specifically the  laser cooled microwave Rb compact clocks whose fractional frequency instability is $ <   8\times 10^{-13}/\sqrt{\tau}$
\cite{Jefferts2018} , and the current global position system (GPS) rubidium atomic frequency standard (RAFS) whose instability is $< 2\times 10^{-12}/\sqrt{\tau}$  
\cite{excelitas} over the same time-frame.  

While laboratory based optical lattice systems provide the highest stability \cite{Mullin1989,Brusch2006,Dzuba2010,Guo2010,Derevianko2011,Ludlow2015}  and progress towards portable lattice clocks is ongoing \cite{Lisdat2017}, a more compact, transportable lattice clock has yet to be fully realized. However, the Optical Rubidium Atomic Frequency Standard (O-RAFS) has potential to meet current needs for compact frequency standards whose stability exceed active hydrogen maser capabilities 
\cite{Maser}, fractional frequency instabilities  $ < 1\times 10^{-13}/\sqrt{\tau}$.  Mitigation of the ac-Stark shift will be an integral component to achieving these ambitious instability goals.  Unlike its lattice clock counterparts which leverage well known ``magic" wavelengths 
\cite{Katori1999,Ye1999}
 to eliminate this light shift, no common light shift mitigation techniques are used for vapour cell clocks.  Because the atoms are sensitive to the average intensity across the vapour, imperfect or unstable optical alignment can also lead to Stark shift fluctuations and clock instabilities.  Moreover, since the atomic vapour is heated to increase signal to noise, the atomic vapour is immersed in blackbody radiation (BBR) 
 \cite{Beloy2012,Safronova2013,Middelmann2012}.  Normally BBR is treated as a dc shift as the blackbody spectrum is far off resonance from the atomic transitions.  However, the $5D_{5/2}$ state in Rb has resonant transitions that are well within the blackbody spectrum requiring calculation as an ac-Stark effect.   Careful calculations and measurements of both the ac- and dc-Stark shifts are required to make predictions of clock performance to determine feasible clock instability goals, and decide whether more complicated Stark shift mitigation techniques are required.

The paper is organized as follows: 
Section \ref{sec:starkshiftcal} details a calculation of the ac-Stark shift at 778.1~nm, Section \ref{sec:starkshiftali} calculates misalignment contributions to the ac-Stark shift, Section \ref{sec:starkshiftmea} describes a measurement of the ac-Stark shift at 778.1~nm,  Section \ref{sec:2color2photon} investigates a two-photon two color approach to excitation of the atom, and Section \ref{sec:bbr} calculates the blackbody radiation shift for the two-photon transition.




\section{Stark Shift at 778.1 nm}

\subsection{Calculation} \label{sec:starkshiftcal}
Although many environmental variables impact the clock instability of O-RAFS, the inherently large ac-Stark shift motivates the most difficult requirements.  Careful calculation and direct measurement of the ac-Stark shift magnitude are pivotal to understanding the overall impact on clock performance.  The ac-Stark shift can be written as
\cite{Grynberg1977},
\begin{equation}\label{eq:SS}
\delta \nu\left(r, z\right) = \frac{\Delta\alpha}{2 c \epsilon_0 h}I\left(r, z \right),
\end{equation}
with $h$ Planck's constant, $c$ the speed of light, $\epsilon_0$ permittivity of free space, $I\left(r, z \right)$ is the laser intensity, $z$ is the optical axis of the beam and $r$ completes the cylindrical coordinate system,  and $\Delta\alpha$, the differential atomic polarizability between the two clock states will need to be calculated. Although the atomic vapour will absorb light from the beam the scattering rate on resonance for the two-photon transition is small, and the laser intensity along the propagation axis can be approximated as constant, $I\left( r,z\right) \approx I\left( r\right)$.
\begin{table*}
\caption{\label{tab:table1} Values for the reduced electric-dipole matrix elements, $\langle J|d|J^{\prime}\rangle$,  taken from 
\cite{Safronova2004}, are presented in a.u., the transition energies are taken from 
\cite{NIST2}, and the Einstein A coefficients are calculated utilizing Equation \ref{eq:EinA}, except as noted.  The chosen sign convention yields negative energies for the $5D_{5/2}\rightarrow 5P_{3/2}$ and $5D_{5/2}\rightarrow 6P_{3/2}$ transitions, indicating that $5D_{5/2}$ is the higher state.}
\begin{ruledtabular}
\begin{tabular}{lccc|lccc}
Transition & $\langle J|d|J^{\prime}\rangle$ & Energy (cm$^{-1}$) & $A_{ki}$ (MCyc/sec) & Transition & $\langle J|d|J^{\prime}\rangle$ & Energy (cm$^{-1}$) & $A_{ki}$ (MCyc/sec)\\
\hline  &&&&&&&\\
$5S_{1/2}\rightarrow 5P_{1/2}$  & 4.233(6) & 12578.950(2) & 36.129(52) \footnote{Lifetime measured in Refs. 
\cite{Volz1996,Simsarian1998,Gutterres2002}} &  $5D_{5/2}\rightarrow 6F_{5/2}$  &1.373(22) \footref{foot:email}&4924.48(2) & 0.076(2) \\
$5S_{1/2}\rightarrow 5P_{3/2}$  & 5.979(9)  &  12816.545(2) & 38.12(13) \footnote{Lifetime measured in Refs. 
\cite{Volz1996,Simsarian1998,Boesten1997,Gutterres2002}} & $5D_{5/2}\rightarrow 7F_{5/2}$  &0.871(38) \footref{foot:email}&5738.23(2) & 0.048(4) \\
$5S_{1/2}\rightarrow 6P_{1/2}$  & 0.3235(9) \footnote{Matrix elements measured in 
\cite{Herold2012}\label{foot:Harold}} & 23715.081(10) & 1.498(8) & $5D_{5/2}\rightarrow 8F_{5/2}$  &0.641(24) \footref{foot:email}&6266.12(2) & 0.034(3)\\
$5S_{1/2}\rightarrow 6P_{3/2}$  &0.5239(8) \footref{foot:Harold} & 23792.591(10) & 1.873(6) & $5D_{5/2}\rightarrow 9F_{5/2}$  &0.513(26) \footref{foot:email}&6629(3) & 0.026(3)\\
$5S_{1/2}\rightarrow 7P_{1/2}$  & 0.101(5) \footnote{Matrix element calculated in Ref. 
\cite{Safronova2004}\label{foot:safronova}} & 27835.02(1) & 0.223(22) & $5D_{5/2}\rightarrow 10F_{5/2}$  &0.440(22) \footref{foot:email}&6889(3) & 0.021(2)\\ 
$5S_{1/2}\rightarrow 7P_{3/2}$  &0.202(10) \footref{foot:safronova}& 27870.11(1) & 0.447(44) & $5D_{5/2}\rightarrow 11F_{5/2}$  &0.450(23) \footref{foot:email}&7080(3) & 0.024(2)\\
$5S_{1/2}\rightarrow 8P_{1/2}$  &0.059(3) \footref{foot:safronova}& 29834.94(1) & 0.094(10) & $5D_{5/2}\rightarrow 12F_{5/2}$  &0.472(71) \footref{foot:email}&7225(3) & 0.028(8)\\ 
$5S_{1/2}\rightarrow 8P_{3/2}$  &0.111(6) \footref{foot:safronova}& 29853.79(1) & 0.166(18) &$5D_{5/2}\rightarrow 13F_{5/2}$  &0.478(72) \footref{foot:email}&7338(3) & 0.030(9) \\ 
$5D_{5/2}\rightarrow 5P_{3/2}$  &1.999(70) \footref{foot:safronova} & -12886.95(5)  &2.89(20) &$5D_{5/2}\rightarrow 4F_{7/2}$  &30.316(64) \footref{foot:email}&1088.59(2) & 0.300(1) \\
$5D_{5/2}\rightarrow 6P_{3/2}$  & 24.621(79) \footref{foot:safronova}& -1910.907(52) &1.428(8) &  $5D_{5/2}\rightarrow 5F_{7/2}$  &11.24(32) \footref{foot:email}&3574.27(2) & 1.461(83)\\
$5D_{5/2}\rightarrow 7P_{3/2}$  &13.82(33) \footref{foot:safronova}& 2166.61(2)&  0.984(47) &$5D_{5/2}\rightarrow 6F_{7/2}$  &6.140(97) \footref{foot:email}&4924.46(2) & 1.140(36)\\
$5D_{5/2}\rightarrow 8P_{3/2}$  &3.292(11) \footref{foot:safronova}& 4150.29(2) &0.392(3) & $5D_{5/2}\rightarrow 7F_{7/2}$  &3.90(17) \footref{foot:email}&5738.22(2)  & 0.726(63)\\
$5D_{5/2}\rightarrow 9P_{3/2}$  &1.691(85) \footnote{Matrix elements derived using the method described in Reference 
\cite{Safronova2004}. \label{foot:email}}& 5266.69(2) & 0.212(21) & $5D_{5/2}\rightarrow 8F_{7/2}$  &2.87(11) \footref{foot:email}&6266.12(2)& 0.512(39)\\
$5D_{5/2}\rightarrow 10P_{3/2}$ &1.099(55) \footref{foot:email}&5957.66(2) & 0.129(13) & $5D_{5/2}\rightarrow 9F_{7/2}$  &2.29(12) \footref{foot:email} &6629(3) & 0.388(41)\\
$5D_{5/2}\rightarrow 11P_{3/2}$ &0.799(40) \footref{foot:email}&6415.02(2) & 0.085(8) & $5D_{5/2}\rightarrow 10F_{7/2}$  &1.968(98) \footref{foot:email} &6889(3) &  0.321(32)\\
$5D_{5/2} \rightarrow 12P_{3/2}$ &0.627(94) \footref{foot:email} &6733.54(6) & 0.061(18) & $5D_{5/2}\rightarrow 11F_{7/2}$  &2.01(10) \footref{foot:email}  &7080(3) &   0.365(4)\\
$5D_{5/2} \rightarrow 13P_{3/2}$ &0.578(87) \footref{foot:email} & 6964.13(6) & 0.057(17) & $5D_{5/2}\rightarrow 12F_{7/2}$  &2.11(32) \footref{foot:email} &7225(3) & 0.425(129)\\
$5D_{5/2}\rightarrow 4F_{5/2}$  &6.779(14)  \footref{foot:email}&1088.62(2) & 0.020(1) & $5D_{5/2}\rightarrow 13F_{7/2}$  &2.14(32) \footref{foot:email} &7338(3) & 0.429(128)\\ 
$5D_{5/2}\rightarrow 5F_{5/2}$  &2.513(72) \footref{foot:email}&3574.29(2) & 0.097(6) &   \\

\end{tabular}
\end{ruledtabular}
\end{table*}


The rank-2 atomic polarizability tensor can be separated into three irreducible components: the scalar (trace), the vector (free symmetric) and the tensor (anti-symmetric) polarizabilites.  The two-photon transition is pumped with linearly polarized light, yielding zero vector shift, and each hyperfine state is addressed uniformly, leaving the atom orientation independent, netting zero tensor shift.  The remaining scalar term is written below in atomic units \cite{Ivan2010,SteckNotes},


\begin{equation}\label{eq:scalepol}
\alpha\left(\omega, J \right) = -\frac{2}{3\left(2 J+1\right)}\sum_{J^{\prime}} \frac{\omega_{J^{\prime},J}|\langle J|d|J^{\prime}\rangle|^2}{\omega_{J^{\prime},J}^2-\omega^2}.
\end{equation}
$\langle J|d|J^{\prime}\rangle$ is the dipole matrix element whose resonant frequency is $\omega_{J^{\prime},J}$; $J$ and $J^{\prime}$ are the associated angular momentum quantum numbers. 


Final calculation of Equation \ref{eq:scalepol} uses the finite basis of B-splines as outlined in Ref. \cite{Johnson1988}.   
Tables \ref{tab:table1} and \ref{tab:continuum} summarize the states and extra considerations included in the polarizability calculation, displaying the transition energy difference, the dipole matrix elements, and the Einstein A coefficients calculated from (unless stated otherwise) \cite{Hilborn2002,Corney1977,Ditchburn1976},
\begin{equation}\label{eq:EinA}
A_{J,J^{\prime}} = \frac{2\omega ^3(e_c a_0 \langle J|d|J^{\prime}\rangle)^2 }{3 \epsilon_0 h c^3(2 J+1)},
\end{equation}
where $e_c$ is the electron charge and $a_0$ is the Bohr radius.  A large number of the parameters listed in Table \ref{tab:table1} originate from Safronova \emph{et al.} 
\cite{Safronova2004}, however, the $5S_{1/2}\rightarrow 5P_{1/2}$ and $5S_{1/2}\rightarrow 5P_{3/2}$ dipole matrix elements are calculated utilizing measured lifetimes
\cite{Volz1996,Simsarian1998,Boesten1997,Gutterres2002}.   The final matrix elements were calculated utilizing the method described in Reference 
\cite{Safronova2004}; these elements only account for a small fraction of the overall differential atomic polarizability.  Uncertanities in the matrix elements were estimated according to References \cite{SAFRONOVA2008, Jiang_2009}.    A majority of the energy levels and uncertainties were obtained from 
\cite{NIST2}. However, the energy levels for the $\left(9-13\right)F$ states were calculated utilizing quantum defect theory(QDT) with a correction accounting for slight discrepancies between observed and predicted energy levels.  QDT generalizes that energy deviations from the Rydberg atom can be written, 
\begin{equation}\label{eq:defect}
E = \frac{A}{(n-d)^2},
\end{equation}
where, E is the energy, n is the principal quantum number and d is the quantum defect.  Equation \ref{eq:defect} can be used to calculate the energies and thus the transition frequencies subtracting the result from the Rb ion limit in 
\cite{Sansonetti2010}.  Necessary elements for the calculation are the Rydberg constant substituted for A from 
\cite{CODATA} and the defects for Rb which are: $S=3.13$,  $P=2.64$, $D=1.35$, and $F=0.016$ 
\cite{Gallagher2011}.   The uncertainty of the energy levels is extrapolated from lower states.

The differential polarizability was calculated for a range of incident wavelengths shown in Figure \ref{fig:twophoton}.   Calculation of the ac-Stark shift at 385.284~THz yields a fractional frequency shift  of   $2.30(4) \times10^{-13}$~/(mW/mm$^2$).  

\begin{table}
\caption{\label{tab:continuum} Contribution to the atomic polarizability from the atomic core and continuum in atomic units.  These numbers were included in the final calculation of ac and dc Stark shift.}
\begin{tabular}{c|c}
\hline
\hline
Contribution & Static Scalar Polarizability\footnote{Polarizability numbers are taken from \cite{Safronova2011}.} \\
\hline
$\alpha_{5S_{1/2}}\left(c\right)$ & 9.1 \\
$\alpha_{5S_{1/2}}\left(tail\right)$ & 1.24 \\
$\alpha_{5D_{5/2}}\left(c\right)$ &  9.0 \\
$\alpha_{5D_{5/2}}\left(> 7P_{3/2}\right)$ & 88(0) \\
$\alpha_{5D_{5/2}}\left(> 6F_{5/2}\right)$ & 13(0) \\
$\alpha_{5D_{5/2}}\left(> 13F_{7/2}\right)$ & 100(0) \\
\hline
\hline
Contribution & Dynamic Scalar Polarizability \\ & at $\lambda=778.1$~nm \\ 
\hline
$\alpha_{5S_{1/2}}\left(c\right)$ & 8.7 \\
$\alpha_{5S_{1/2}}\left(tail\right)$ & 0.5 \\
$\alpha_{5D_{5/2}}\left(c\right)$ &  9.0 \\
$\alpha_{5D_{5/2}}\left(> 13P_{3/2}\right)$ & -4(74) \\
$\alpha_{5D_{5/2}}\left(> 13F_{5/2}\right)$ & -1.8(5) \\
$\alpha_{5D_{5/2}}\left(> 13F_{7/2}\right)$ & -33(110) \\
\end{tabular}
\end{table}

%


Equation \ref{eq:SS} shows an ac-Stark shift dependence on local intensity of the laser electric field.  The atomic vapour effectively samples the laser intensity distribution, and the fluorescence spectrum is shifted on average by
\begin{equation}\label{eq:averageStark}
\overline{\delta \nu} = \frac{\Delta\alpha}{2 c \epsilon_0 h}\overline{I_{tot}},
\end{equation}
$\overline{I_{tot}}$ is the spatial weighted average intensity,
\begin{equation} \label{eq:IWA}
\overline{I_{tot}} = \displaystyle \frac{\int I_{tot}I_1 I_2 dV}{\int I_1  I_2  dV},
\end{equation}
with the total intensity as the sum of the two beam profiles $I_{tot} = I_1 + I_2$.  The weighting function $I_1I_2$ is utilized as a measure of the intensity in the two-photon region, the beam overlap in the vapour cell, the main contributor to the ac-Stark shift.    Equation \ref{eq:averageStark} details that variations in laser intensity will cause a clock shift,  and these can arise from laser power fluctuations as well as slight optical alignment variations.    


\begin{figure}
\includegraphics[width=0.47\textwidth]{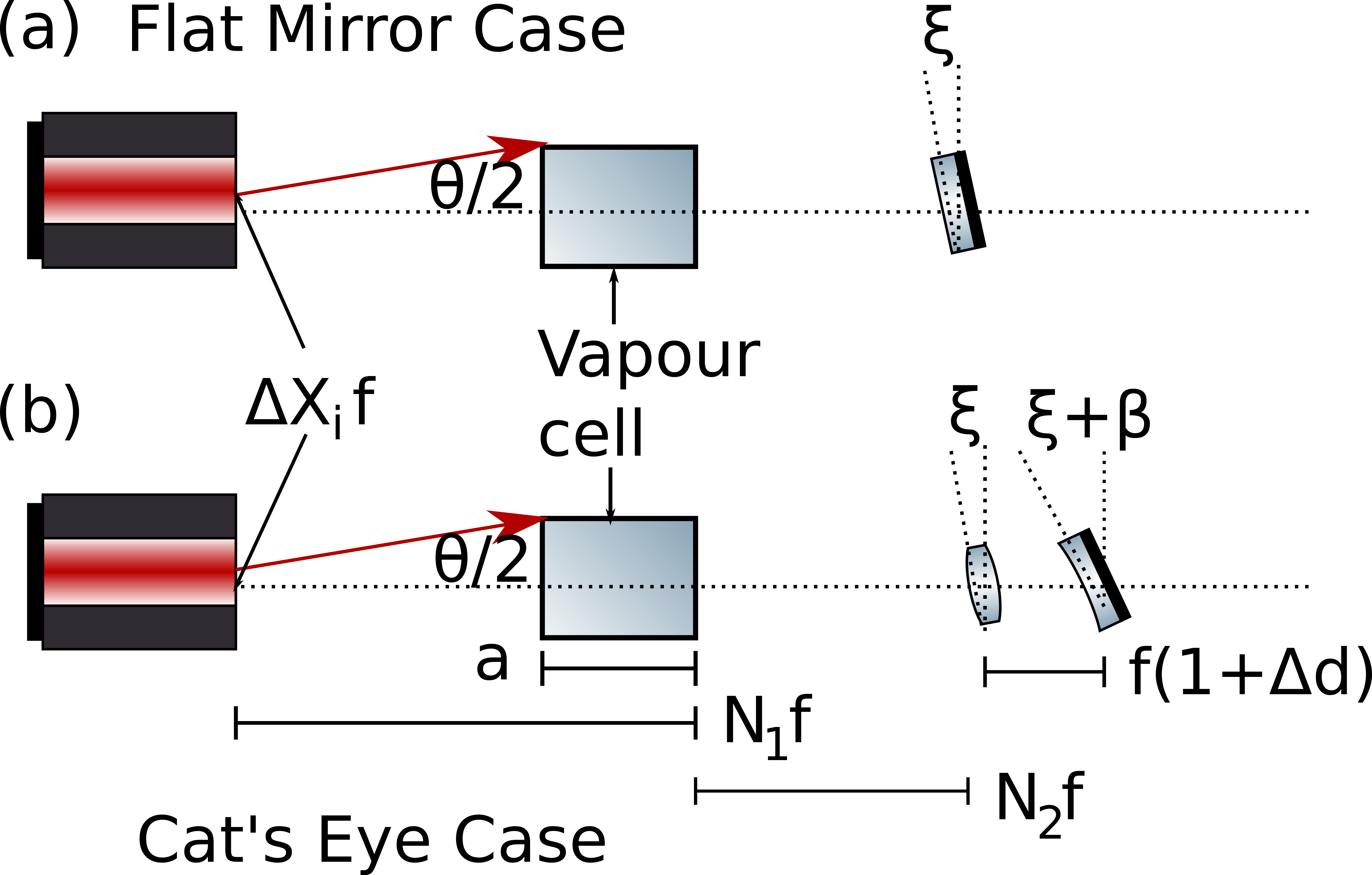}
\caption{\label{fig:cats} Simple geometric alignment diagram for a flat mirror (a) and a cat's eye (b).  In each case all distances were written in terms of the focal length of the lens utilized in the cat's eye: $N_1$, $N_2$, $\Delta d$, $\Delta$f $\Delta X_i$ are unit-less parameters. $\Delta X_if$ - displacement of the initial Gaussian beam from the optical axis, $\theta$ - angular misalignment of the initial beam, $\xi$ - angular misalignment of the retro-reflecting optic, $\beta$ - angular misalignment of the mirror and lens in the cat's eye optic, $f\Delta d$ - displacement from optimal placement of the mirror in the cat's eye optic, $N_1f$ - distance from the fiber launcher to the rear of the vapour cell, $N_2f$ - distance from the vapour cell to the retro-reflector, $a$ -vapour cell length and $f\Delta f$ - absolute deviation of the cat's eye mirror from optimal radius of curvature.  }
\end{figure}

\subsection{Alignment Shift} \label{sec:starkshiftali}

Typical optic mounts allow slight tip/tilt adjustments which are required for control of the beam.  In the O-RAFS system, variations in alignment affect the total intensity of the two-photon excitation causing an ac-Stark instability.    This section examines slight angular variations of the laser light emitted from the fiber launcher, as well as slight variation of the retro-reflector, and calculates the variation of $\bar{I_{tot}}$.  Figure \ref{fig:cats} presents what a small angular misalignment would look like for both a flat mirror (a) and a cat's eye retro-reflector (b).  These two separate retro-reflectors have been successfully employed in the experimental apparatus. The optimal retro-reflector can be determined through calculation of Equation \ref{eq:IWA}.

Normally, the intensity of a Gaussian beam after interaction with a series of optical components can be calculated using the ABCD or M matrices and the complex beam parameter $q$
\cite{Siegman1986, SteckNotes}.  However, the M matrices require that the optical elements are placed normal to the optical axis, which is not true for a general case. Instead, we use the extended ray trace matrices  
\cite{Siegman1986} (see Table \ref{tab:raymatrix}),
\begin{equation}
\begin{pmatrix}
   A & B & \delta \\  
   C & D & \gamma \\  
   0 &   0 &  1    
\end{pmatrix},
\end{equation}
where $\delta$ is a displacement from the optical axis, $\gamma$  is a rotation of the optic from normal incidence, and the ABCD elements are unchanged.

 \begin{table*}
\caption{\label{tab:raymatrix} The extended ray trace matrices used to calculate misalignment effects from the cat's eye optic}
\begin{tabular}{l|l}
\hline \\

Freespace propagation of distance $N$ &
$
  \begin{pmatrix}
    1&N&0\\
    0&1&0\\
    0&0&1
  \end{pmatrix}
$\\
Thin lens with focal length f at angle $\xi$ &
$
  \begin{pmatrix}
    1&0&0\\
    -1/f&1&\xi\\
    0&&1
  \end{pmatrix}
$\\ 
Concave mirror, radius $f(1+\Delta f)$, at angle $\xi+\beta$ &
$
  \begin{pmatrix}
    1&0&0\\
    2/f(1+\Delta f)&1&\xi+\beta\\
    0&0&1
  \end{pmatrix}
$\\ 
Flat mirror, at angle $\xi$ &
$
  \begin{pmatrix}
    1&0&0\\
    0&1&\xi\\
    0&0&1
  \end{pmatrix}
$\\ 
\hline
\end{tabular}
\end{table*}

\begin{figure}
\begin{center}
	\includegraphics[width=0.47\textwidth]{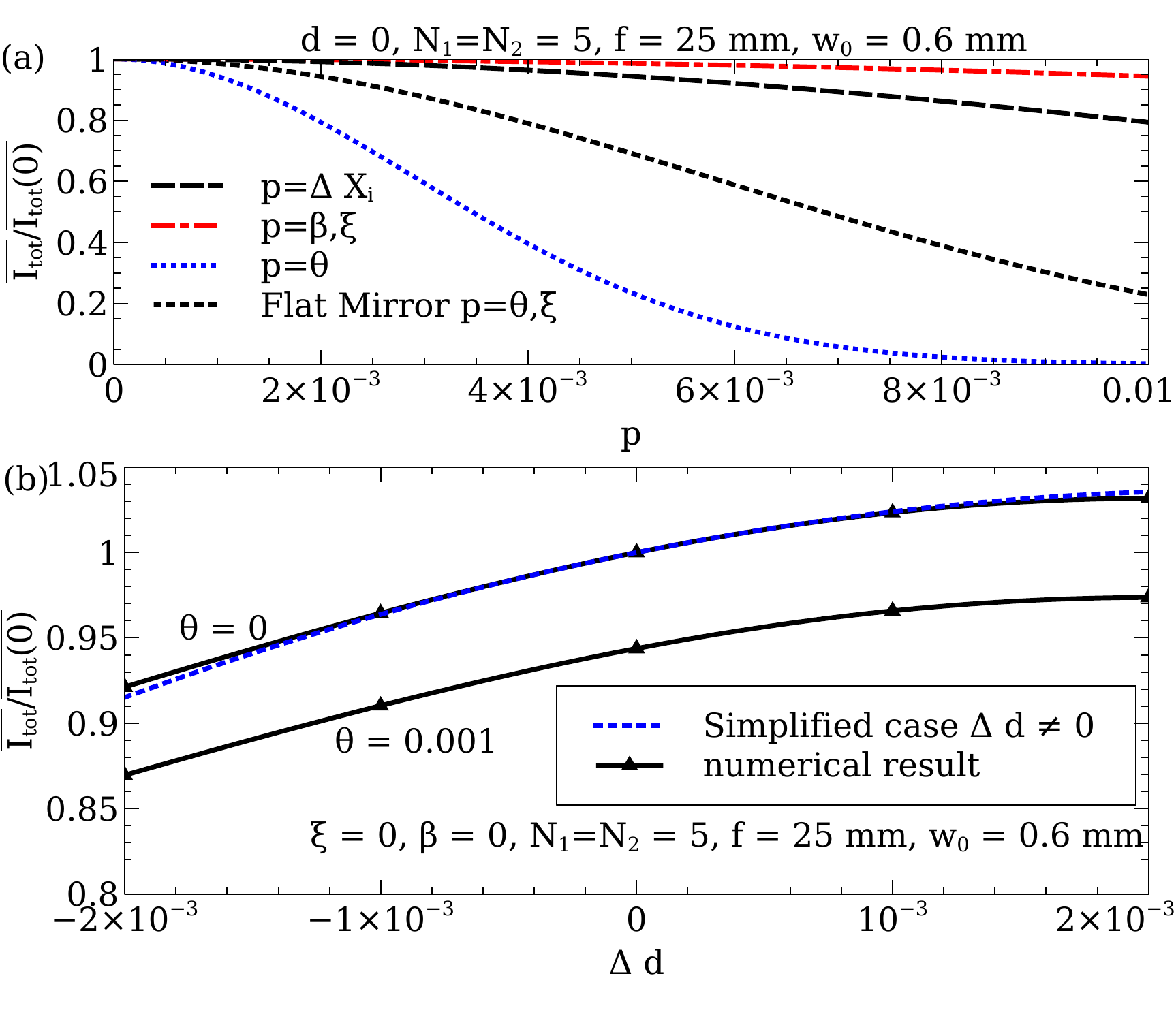}
	\end{center}
	\caption{\label{fig:d0alignment} Shown in (a) is the simplified case where $\Delta d=0$.  Clearly, the fiber launcher angle is more sensitive than the other misalignment variables in the cat's eye case and either angular misalignment in the flat mirror case.  Shown in (b) is the numerically calculated average intensity as a function of $\Delta d$ for incident angular misalignment of $\theta=0$ and $\theta=0.001$ for $\Delta d\neq0$  as well as the simplified case derived in the appendix.  The numerical result with $\theta=0$ and the simplified case in Equation \ref{eq:simple} agree over a half percent change in cat's eye optic displacement (the dotted blue line is obscured by the solid black line).}

\end{figure}

In the flat mirror case, the Gaussian beam originating from the fiber launcher can be written as
\begin{equation}
I_{incident}=I_0 e^{-2(x^2+y^2)/w_0^2},
\end{equation}
with $w_0$ the $1/e^2$ intensity radius.  The extended M matrix for the beam as it re-enters the vapour cell after reflection off the mirror is 
\begin{equation} \label{eq:flatmirrorretro}
  M = \begin{pmatrix}
    1&2N_2f&N_2f\xi\\
    0&1&\xi\\
    0&0&1
  \end{pmatrix}.
\end{equation}
The complex beam paramter $q$ \cite{Siegman1986,SteckNotes},
\begin{equation} 
q=q_1+iq_2=\frac{Aw_0^2\pi i/\lambda  +B}{Cw_0^2\pi i/\lambda  +D},
\end{equation}
can be used to to determine the retro reflected beam profile as it re-enters the cell. Substituting the parameters from Equation  \ref{eq:flatmirrorretro} yields,   
\begin{equation}
q=w_0^2\pi i/\lambda  +2N_2f,
\end{equation}
where $\lambda$ is the wavelength.  Calculating the retro-reflected beam from these q-parameters and using the approximation that the free space propagation lengths are less than the Rayleigh length and thus $(q_1/q_2)<<1$,  yields,
\begin{equation}
I_{retro}=I_0 e^{-2(x^2+y^2+2x(N_2f+z)(\xi-\theta)+(N_2f+z)^2(\xi-\theta)^2)/w_0^2},
\end{equation}
for the retro-reflected beam.  Equation \ref{eq:IWA} is calculated and the normalized weighted average, $\overline{I}$, is given by, 
\begin{equation}
\overline{I}=\frac{\overline{I_{tot}}}{\overline{I_{tot}}(\xi=\theta/2)} = \frac{\sqrt{3}}{2}\frac{\erf\left(2(y_1+y_2)/\sqrt{3}\right)-\erf\left(2y_2/\sqrt{3}\right)}{\erf\left(y_1+y_2\right)-\erf\left(y_2\right)},
\end{equation}
where $y_1=a\left(\xi-\theta\right)/\omega_0$ and $y_2=Mf\left(\xi-\theta\right)/\omega_0$.  This result is plotted in Figure  \ref{fig:d0alignment} for the experimental set-up described in 
\cite{Martin2018}.
The M matrix formulation leverages the small angle approximation where $\xi,\theta <<1$.  If the higher order terms are ignored, the weighted average becomes 
\begin{equation} \label{eq:seriesTWI}
\overline{I} \approx 1-\frac{\left(a^2+3aN_2f+3N_2^2f^2\right)\left(\xi-\theta\right)^2}{9 \omega_0^2}, 
\end{equation}
which has been expanded to second order in $\theta$, $\xi$ and their cross terms.

The final desired quantity is the sensitivity of the average weighted total intensity to angular misalignment in each optic, or the derivative of the intensity w.r.t. the angular variable.  Taking the derivative of Equation \ref{eq:seriesTWI} w.r.t. either $\theta$ or $\xi$ yield the same result, 
\begin{equation}
\frac{d\overline{I} }{d\theta} \approx \frac{-2\left(a^2+3aN_2f+3N_2^2f^2\right)\theta}{9 \omega_0^2}.
\end{equation}


The cat's eye calculation is more challenging.  A cat's eye retro-reflector consists of a convex lens with a focal length f, and a concave mirror with a radius of curvature f, placed at the focus of the lens.  An ideal cat's eye provides an anti-parallel retro-reflected beam 
\cite{Snyder1975}, whereas misalignment in the fiber launcher will yield non-parallel beams in the case of a flat mirror reflector.     Although the incident beam can be recycled and the q-parameter formulation is the same, the M matrix becomes more complex.  Various misalignment mechanisms in the cat's eye optic calculation, shown in Figure \ref{fig:cats}, include small deviations of the location of the mirror (now located at $f(1+\Delta d)$) the radius of curvature of the mirror ($f(1+\Delta f)$), and also allowing the lens and mirror to have a misalignment angle $\beta$ between them.  The following assumptions are enforced: $\xi << 1$, $\theta << 1$, $\Delta d << 1$, $\Delta f << 1$, $\Delta X_i << 1$, $\beta << 1$, and the initial beam is collimated.  Three special cases were examined, and details on each special case can be found in the appendix:
\begin{itemize}
\item[1.] Analytic derivation with $\Delta d=0$.
\item[2.] One specific numeric case with $\Delta d\neq0$.
\item[3.] Analytic derivation with $\Delta d\neq0$, $\Delta X_i=0$, $\beta= 0$  and $\Delta f = 0$
\end{itemize}
Figure \ref{fig:d0alignment} shows the variation in average intensity as a function of $\Delta X_i$, $\beta$, $\xi$, and $\theta$ with all other misalignment variables set explicitly to zero.  The fiber launcher angular sensitivity dominates in this simple case.  As in the flat mirror case the sensitivity of the average intensity w.r.t. angular misalignments of the fiber launcher and retro-reflector is the desired quantity.  In the cat's eye case these solutions differ, 
\begin{equation} \label{eq:cats_alignment_1}
\frac{d\overline{I} }{d\theta} \approx \frac{-2f^2\left(N_1+N_2\right)\left(\xi+\left(N_1+N_2\right)\theta\right)}{3 \omega_0^2}
\end{equation}
\begin{equation}\label{eq:cats_alignment_2}
\frac{d\overline{I} }{d\xi} \approx \frac{-2f^2\left(\xi+\left(N_1+N_2\right)\theta\right)}{3 \omega_0^2} \\
\end{equation}
where $\Delta f =0$, $\beta = 0$ , and $\Delta X_i = 0$. It is apparent that Equation \ref{eq:cats_alignment_1} is larger than Equation \ref{eq:cats_alignment_2} by a factor of $N_1+N_2$.  By design, the cat's eye suppresses angular motion of the retro-reflector optic over the flat mirror, but at the expense of sensitivity to angular misalignment of the fiber launcher.  

The choice of retro-reflector will depend on environmental conditions.  For a fully dynamic environment, the flat mirror case might be the best choice, however, if the collimator can be rigidly mounted, the cat's eye retro-reflector is ideal for reducing alignment variations.   Moreover, reduction of alignment sensitivity can be achieved by reducing the laser intensity or increasing the beam waist.

\begin{figure*}
\includegraphics[width=\textwidth]{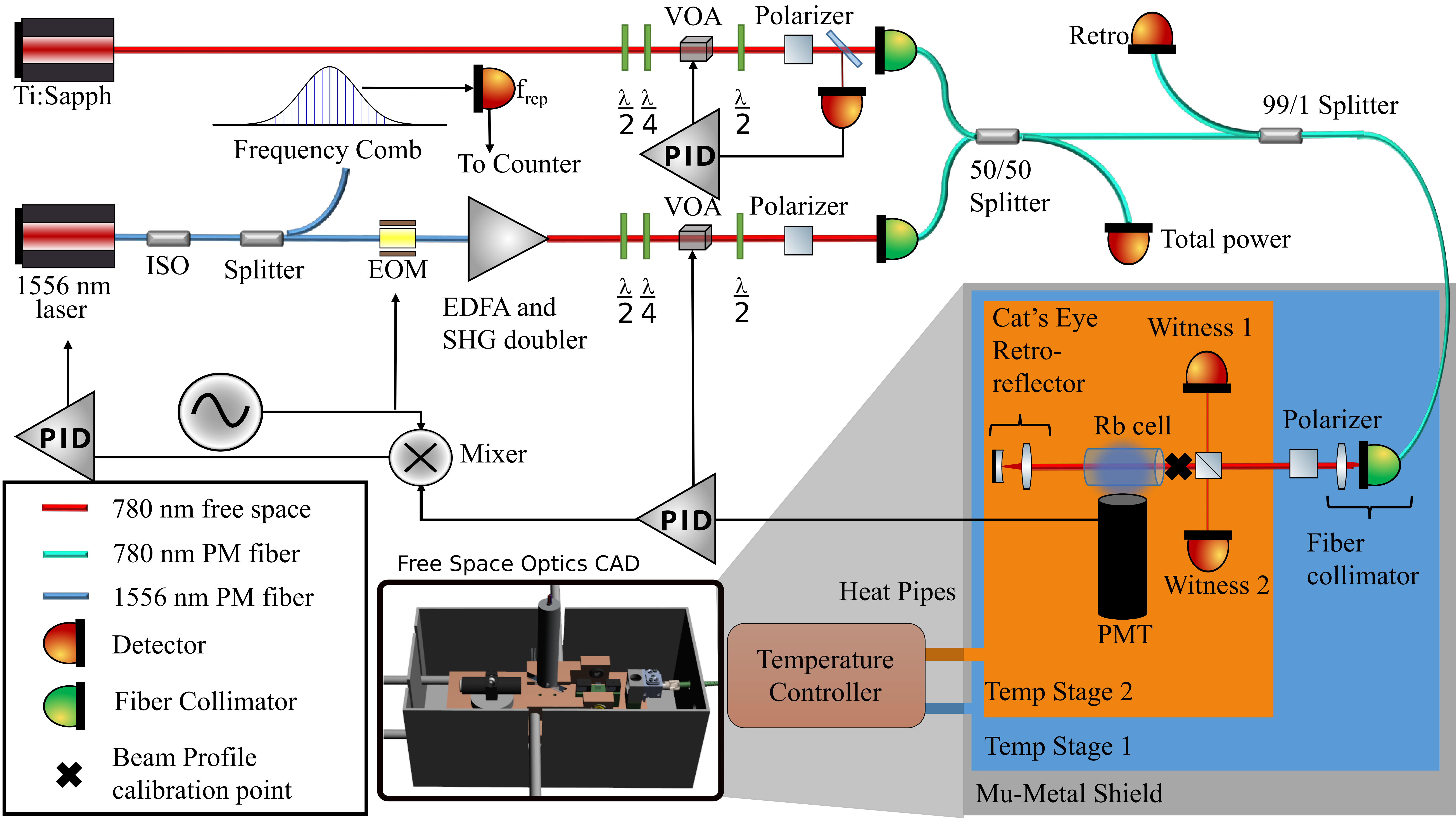}
\caption{\label{fig:quantel} An optical and simplified electrical schematic of the Rb two-photon frequency standard as described in the text. Also shown is a computer animated drawing of the rubidium atomic reference.  EOM - electro-optic modulator; PMT - photomultiplier tube;  VOA - variable optical attenuator; ISO - optical isolator; EDFA - erbium doped fiber amplifier; SHG - second harmonic generator;  PID - proportional integral differential lock mechanism.}
\end{figure*}

\subsection{Measurement}\label{sec:starkshiftmea}

Measuring the ac-Stark shift is important to confirm the accuracy of the theoretical result. An experiment was designed specifically to measure the effects of the ac-Stark shift on the output clock frequency.  Two lasers were utilized in the Stark shift measurement (Figure \ref{fig:quantel}): a Ti:Sapph laser tuned slightly away from the two-photon resonance to 385287.8~GHz, which is far enough from the resonant two-photon excitation frequency to introduce no measurable vapour excitation, and the clock laser tuned to be on resonance with the two-photon transition, described in \cite{Martin2018}.   After amplification and subsequent second harmonic generation of 778.1~nm, the clock laser and the Ti:Sapph laser separately pass through a half wave-plate followed by a quarter wave-plate.  The remaining light in each beam is subsequently sent through a variable optical attenuator (VOA), which is used for laser power stabilization. Each beam then passes through another half wave-plate and polarizer, properly aligning the polarization to fiber couple each beam into two separate arms of a polarization maintaining (PM) $2\times2$ 50:50 fiber splitter.  A portion of the Ti:Sapph light is sampled before fiber coupling.  This signal is used to feedback to the Ti:Sapph VOA and stabilize the optical power coupled into the beam splitter.  One arm of the splitter is sent to a detector used for independent power measurements. The second arm of the splitter is sent to the vapour cell assembly through a $2\times2$ 99:1 PM fiber splitter.  Up to 30~mW of Ti:Sapph light and 30 mW of clock laser light were delivered to the vapour cell assembly.

The vapour cell assembly is enclosed in 5~mm thick single layer mu-metal magnetic shield to reduce Zeeman shifts and broadening.   The atomic vapour is regulated to a constant temperature, ideally 100~$^{\circ}$C.  The final vapour cell operational temperature is achieved though use of a dual stage temperature apparatus.  The first temperature state is regulated to $60$~$^{\circ}$C, Temp stage 1 in Figure~\ref{fig:quantel}, providing a stable reference for the final stage regulated to 100~$^{\circ}$C, Temp stage 2 in Figure~\ref{fig:quantel}.  The vapor cell, which is a rectangular prism with dimensions 5~mm $\times$ 5~mm $\times$ 25~mm, containing $>99\%$ isotopically enriched $^{87}$Rb, is placed such that it has a $1~$K thermal gradient along its length, forcing the cell's cold spot on the pinched-off fill tube of the borosilicate glass cell.  The vapor cell is oriented at Brewster's angle with respect to the incident laser beam to reduce stray reflections. The vapour cell assembly is further described in 
\cite{Martin2018}.

\begin{figure}
\includegraphics[width=0.47\textwidth]{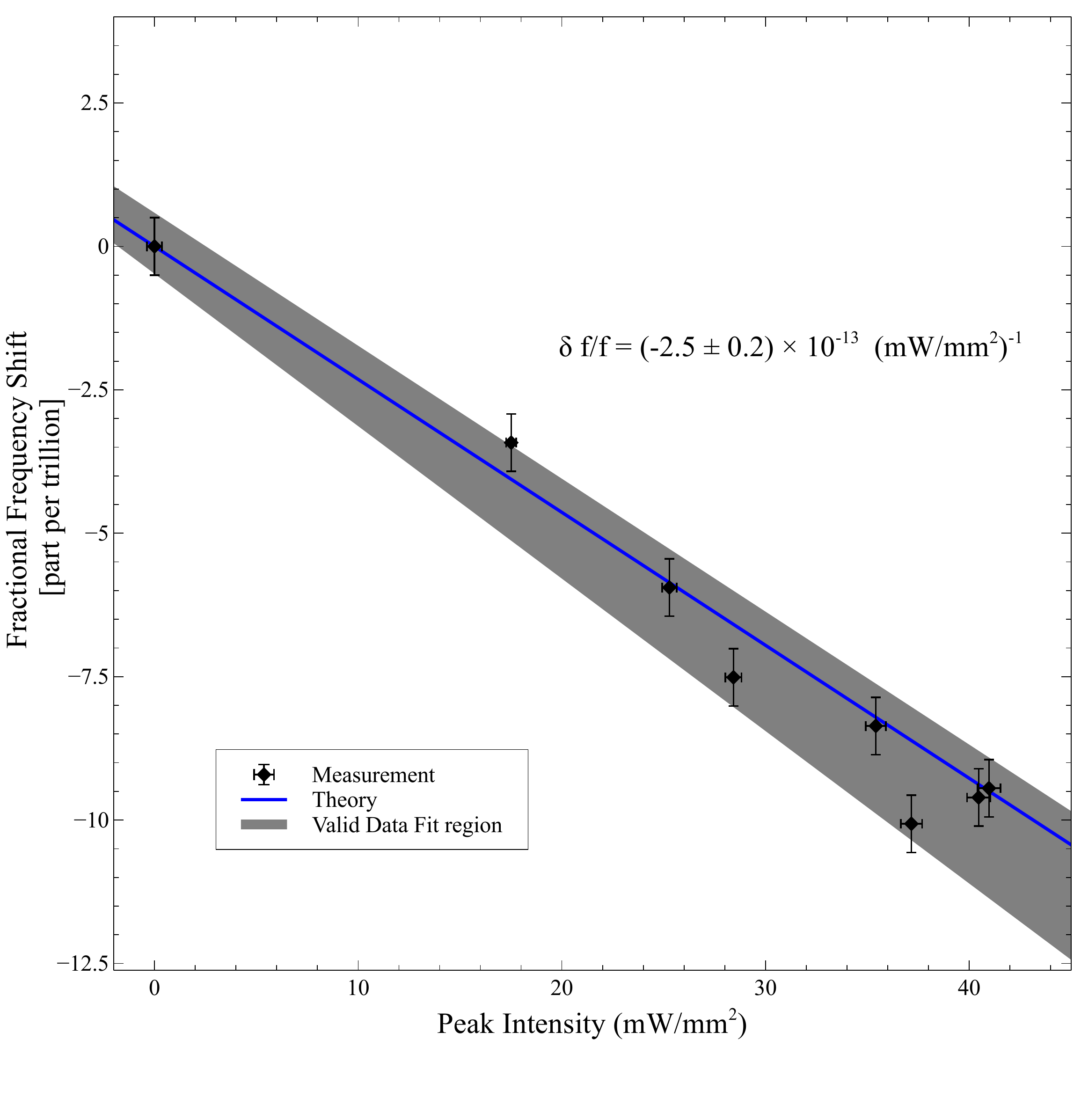}
\caption{\label{fig:light_shift} Experimentally measured 778~nm ac-Stark shift for a (0.66 $\pm$ 0.05) mm beam plotted in terms of peak intensity.  An orthogonal distance regression considering errorbars in both x and y coordinates determined a frequency shift of $-2.5(2)\times10^{-13}$~(mW/mm$^{2}$)$^{-1}$ with a reduced $\chi^2$ of 1.57.  The calculated ac-Stark shift is also shown in blue.}
\end{figure}

The two witness photodiodes, a Thorlabs SM05PD1A, labeled  witness 1 in the diagram, and a Thorlabs PDA36A, labeled by the total power detector on in the diagram were independently calibrated to both a Thorlabs PM160 and an Ophir PD300-TP hand held silicon power meter at the "X" marked in Figure \ref{fig:quantel}.  Differences in the calibrations are included in the combined statistical and systematic errorbars of Figure \ref{fig:light_shift}.

After the clock laser is stabilized to be on resonance, the Ti:Sapph laser power is varied using the VOA in its optical train.  The frequency shifts are measured and averaged over 100 seconds and are reported in Figure \ref{fig:light_shift}.  The data was fit with a orthogonal distance regression (ODR) algorithm which weights the error bars in both the x and y coordinates, yielding a fractional frequency fit of -2.5(2)$\times 10^{-13}$ (mW/mm$^2$)$^{-1}$ with a reduced $\chi^2$ of 1.57.  A gray shaded region shows associated error with the fit.  The theoretical value is plotted on the same curve and shows good agreement with the experimentally measured values.



\section{Two Color}\label{sec:2color2photon}
Reduction of the ac-Stark shift would be a powerful tool to reach the final long-term instability goal of $1\times10^{-15}$ at one day.   Alternative to the outlined approach in 
\cite{Martin2018}, two lasers of different frequencies could be utilized to excite the atom in a two-color approach described in 
\cite{Perrella2013,Gerginov2018,Perrella:13}.    
\begin{figure}
\includegraphics[width=0.47\textwidth]{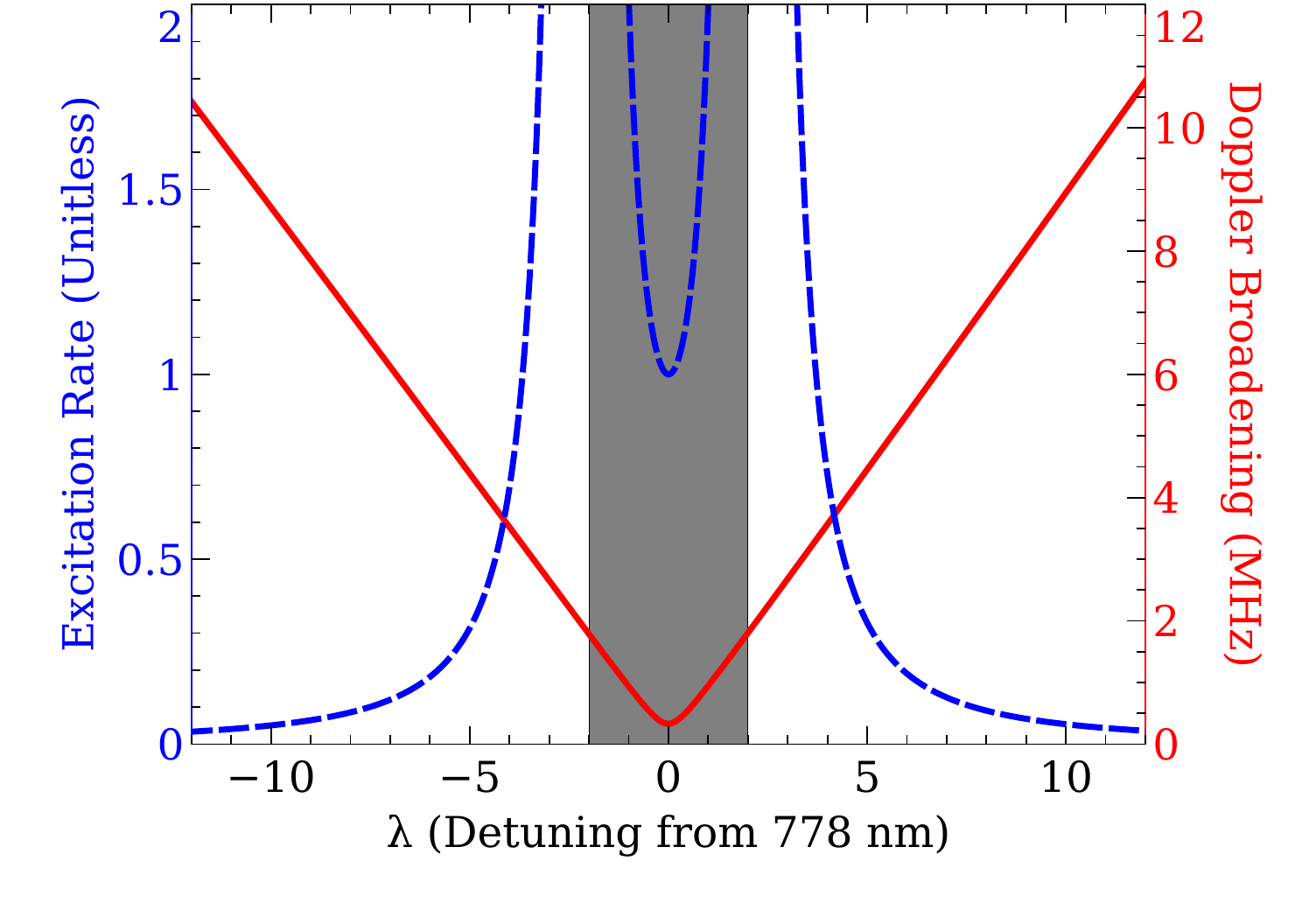}
\caption{\label{fig:twocolor} Shown above is the calculated Doppler broadened peak (red solid line) and the normalized excitation rate (blue dashed line) for a two photon transition where the second photon is constrained to keep the total transition on resonance. The grey shaded region shows a choice of photon pair that will not lead to ac-Stark cancellation. } 
\end{figure} 
However, this approach leads to residual Doppler broadening, because in the atomic frame the $\vv{\bm{k}}$ vectors of the excitation photons no longer match. The increased spectral width requires a higher excitation rate to achieve similar clock performance motivating the choice to operate this two-color scheme near resonance (see Figure \ref{fig:twocolor}).  Short term stability has been shown to remain unchanged in this design.  However, clock stability on timescales exceeding a few hours has not yet been measured.  Instabilities on longer timescales can often be driven by technical noise, i.e. lock-point errors, reference voltage drifts, detector responsivity drift, etc. Broadening the transition could make control of these noise sources more difficult.  

An alternative two-color approach introduces a second laser (off-resonance from the virtual and intermediate states) to the degenerate two photon Doppler free experiment.  The wavelength of this mitigation laser would be chosen such that the differential polarizability sign is opposite that of a 778.1~nm photon.  Normally, the clock laser intensity is measured and stabilized to a voltage reference.  This scheme is sensitive to stabilities of the detector and voltage reference as well as drift in each device.  The two-color approach would allow for the laser intensities to be stabilized with respect to each other reducing requirements on detector sensitivities and drift as well as eliminating a need for a precise voltage reference.  Unfortunately, frequency drifts in the mitigation laser would cause  variations in the ideal ratio, imposing requirements on the frequency stability.  Volumetric Bragg grating stabilized lasers developed for Raman spectroscopy experiments 
\cite{Rosser2008,Knorr2010,Vorobiev2008,Martel2010} offer potential options for mitigation lasers.  Table \ref{tab:mitagating_laser} displays three possible mitigating wavelengths, the clock laser and calculated shifts.  For a system whose purpose is to minimize the required operational power while maintaining high short term clock stability the mitigation laser at 785~nm is the preferred choice.  

Use of a mitigation laser could ease the challenge of power stabilizing the probe laser.  O-RAFS has already demonstrated that clock laser power can be stabilized to 0.1\% 
\cite{Martin2018}.  Current requirements on laser stability however, require absolute laser stabilization to 0.01\%.  If the cancellation can be maintained at a 0.1\% level stabilization of the ac-Stark shift can be maintained at 0.01\% allowing for final clock stability of $1\times 10^{-15}$ to be achieved.

\begin{table}
\caption{\label{tab:mitagating_laser} Proposed wavelengths, associated shifts and required power (mW)  per mW of 778~nm for a ac-Stark mitigation laser.}
\begin{tabular}{l|cr}
Wavelength & ac-Stark shift & Power multiplier \\
(nm) & (Hz/(mW/mm$^2$)) & $\times$P$_{778}$\\
\hline \\
1556.2 & -16.5 & 10.8\\
808 & -30.9 & 5.7\\
785 & -62.5 & 2.9\\
778 & 178.5 & X\\
\end{tabular}
\end{table}

\section{Blackbody Radiation Shift}\label{sec:bbr}

Plank's Law describes the electromagnetic radiation emanating from an object at temperature $T$.  The time averaged intensity of the radiated field can expressed, 
\cite{Mullin1989}, 
\begin{equation} \label{eq:Mullin}
\langle E\left(\omega\right)\rangle^2 = \frac{\hbar}{\pi^2 \epsilon_0 c^3}\frac{\omega^3}{e^{\hbar\omega/k_BT}-1}.
\end{equation}
An atom in a bath of electromagnetic radiation will experience an ac-Stark shift.   Equation \ref{eq:Mullin} describes an electromagnetic field whose intensity is dependent on the source temperature.  Thus, fluctuating temperatures of a source that is radiatively coupled to the atoms can cause a clock shift via temperature driven ac-Stark interactions.  The shift arising from blackbody radiation (BBR) can be calculated by, 
\begin{equation}\label{eq:BBRint}
\delta\nu=\frac{1}{2 h}\displaystyle\int_{0}^{\infty}\Delta \alpha (\omega)E\left(\omega\right)^2  d\omega, 
\end{equation}
where $\Delta\alpha(\omega)=\alpha_e-\alpha_g$.  Oftentimes, the resonance frequencies between atomic states connected to the ground and excited states involved in the clock transition are far from the blackbody spectrum.  In this case the blackbody spectrum can be treated as a static polarizability field and Equation \ref{eq:BBRint} can be simplified to be,
\begin{equation}\label{eq:BBRintsimple}
\delta\nu=\frac{\Delta\alpha}{2 h}\displaystyle\int_{0}^{\infty}E\left(\omega\right)^2  d\omega \approx \frac{\Delta\alpha}{2h}\left(8.3~\frac{V}{cm}\right)^2\left(T/300~K\right)^4.
\end{equation}
Systems that require higher precision include a small dynamic contribution, $\eta$, to account for frequency dependence \cite{Beloy2012,Safronova2013,Middelmann2012}.  However, for the Rb two-photon transition the wide-band BBR spectrum has significant overlap with the  $5D_{5/2}\rightarrow 4F_{7/2,5/2}$ transitions at operational temperatures making it necessary to fully integrate Equation \ref{eq:BBRint}.

This integral was calculated numerically two separate ways.   First, the polarizability was calculated using Equation \ref{eq:scalepol} and the integral was performed using the Cauchy's principle value.    The second method 
\cite{SteckNotes} introduced  the decay rate , $\Gamma_j$, of each transition found in 
\cite{Marek_1980,Safronova2011},
\begin{equation}
 \alpha(\omega,J) = -\frac{2}{3(2J+1)\hbar}\displaystyle\sum_{J^{\prime}}\frac{\omega_{J^{\prime},J}|\langle J|d|J^{\prime}\rangle|^2 (\omega_{J^{\prime},J}^2-\omega^2)}{(\omega_{J^{\prime},J}^2-\omega^2)^2+\Gamma_J^2 \omega^2}.
\end{equation}
  When the polarizability is written this way the function no longer diverges in the resonant cases.  Equation \ref{eq:BBRint} was then calculated with a deterministic adaptive integration technique.  BBR shifts were examined under nominal operational conditions, specifically looking at the shifts produced in the temperature range of 300~K to 500~K.  The fractional difference for each of these calculations for both the excited state and ground state were less then $1\times 10^{-5}$.   
\begin{figure}
\includegraphics[width=0.47\textwidth]{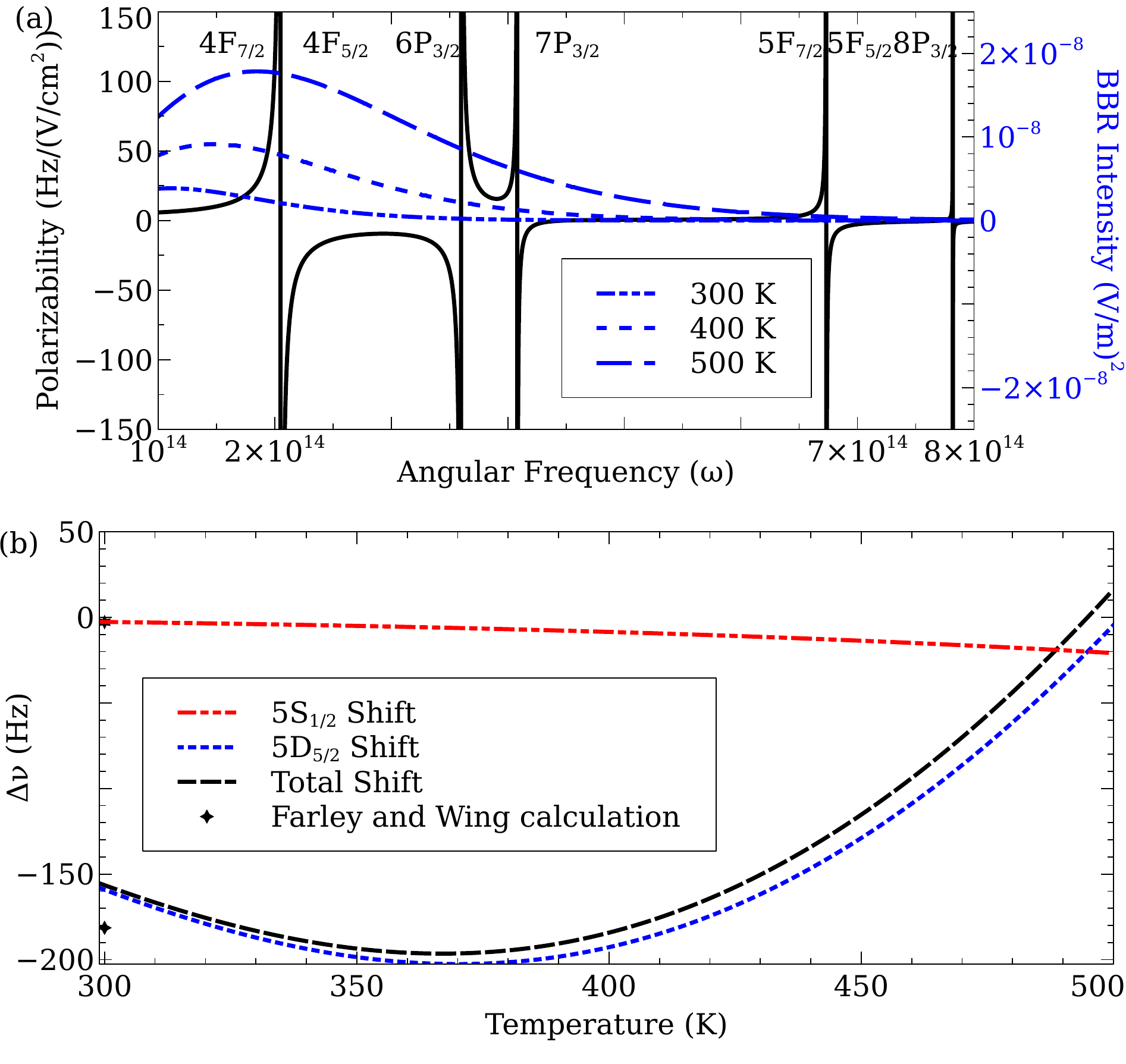}
\caption{\label{fig:bbr}  Plotted in (a) is the blackbody radiation intensity (blue) as a function of angular frequency for and environment at 300~K (dashed-dotted line), 400~K (dotted line), and 500~K (dashed line).  Also shown in (a) is the differential polarizability (black solid line) as a function of angular frequency.  Along the top of (a) are labeled the intermediate states that connect to $5D_{5/2}$ and whose spectrum overlaps that of a blackbody at operational temperature. (b) The state shifts and the total BBR shift for the two-photon transition are plotted as a function of temperature.} 
\end{figure} 

The resultant BBR shift of the $5S_{1/2}$ ground state, $-$2.68(2)~Hz, whose polarizabilities are far off resonance for the examined temperature range, yields a result that was consistent with the Farley \emph{et al.} 
\cite{Farley1981} calculations of $-2.789$~Hz at 300~K.  The shift also is consistent with a $T^4$ temperature dependence and a static polarizability approximation.  More interesting were the results from integration over the excited state polarizabilities.  At $300$~K our value of $-$158.4(12)~Hz differed from Farley \emph{et al.} calculation of $-181.4$~Hz, likely due to differences in polarizability values.  Regardless of this difference, the $5D_{5/2}$ state has resonant polarizablities in the temperature range of interest.  Not only was the calculated ac-Stark shift no longer monotonic, the differential polarizability changed sign (see Figure \ref{fig:bbr}).   This result is not consistent with either a $T^4$ behavior or with the static polarizability approximation.  
The calculation yields two interesting ``magic" temperatures.  Around 495.9(27)~K, $\delta\nu=0$ and around 368.1(14)~K, the BBR shift is insensitive to changes in environmental temperature.  Clock operation hoping to achieve greater stability could operate around 368~K to suppress environmental temperature dependence.  However, to benefit from this reduced temperature sensitivity, the more significant temperature shift arising from Rb collisional effects would also require mitigation 
\cite{Martin2018,Zameroski2014}.



         


\section{dc-Stark}


The presence of unknown electric charges on dielectric surfaces in the vicinity of the atomic sample causes a DC Stark shift and could, at least in principle, cause a clock instability if the charge was to slowly migrate. In fact, the buildup of stray charge inside of a vacuum chamber on a mirror with a piezoelectric transducer was shown to be 
detrimental to the overall performance in an optical lattice clock \cite{Lodewyck2011}. We calculate the DC polarizability for the Rb 2-photon transition from Equations \ref{eq:SS} and \ref{eq:scalepol}, setting $\omega = 0$. The polarizability of the $5D_{5/2}$ state dominates due to the presence of low-lying resonances, and we find the induced Stark shift to be 4.27(4) Hz/(V/cm)$^2$ by summing over the matrix elements in Table \ref{tab:table1} and inclusion of the core and continuum polarizabilites found in Table \ref{tab:continuum}. When probed with a 778 nm laser, the 2-photon frequency standard displays a fractional sensitivity of 5.55(5) $\times$ 10$^{-15}$ /(V/cm)$^2$. 


Thus, in principle, a small electric field on the order of 1~V/cm with a slow time wander could cause a significant long-term stability for this frequency standard. However, since the glass vapor cell is embedded in a block of copper for thermal control, we expect this effect to be minimal. The inclusion of a UV LED to remove the charge  \cite{Lodewyck2011} is one simple way to rule out this possibility in future investigations of long-term instabilities. 

\section{Conclusions}
We presented a calculated ac-Stark shift of $2.30(4)\times10^{-13}$(mW/mm$^2$)$^{-1}$ in good agreement with the measured value of $2.5(2)\times10^{-13}$(mW/mm$^2$)$^{-1}$.  Careful examination of alignment shows that the cat's eye retro-reflector helps reduce sensitivity of the average intensity to variations in the retro-reflecting optic over the flat mirror reflector.  However, the cat's eye retro-reflector increases the sensitivity of the average intensity to angular misalignments of the fiber launcher, and introduces another very sensitive misalignment variable, the distance between the lens and mirror in the cat's eye optic.  While effort can be placed to reduce the dynamic response of the displacement $\Delta d$, care must be taken to ensure that the average intensity signal is maximized during initial alignment. For a dynamic system, however,  the flat mirror retro-reflector might be the best choice to reduce complexity and sensitivity to motion.   Ultimately, any effort to further reduce the overall ac-Stark shift will also reduce Stark shift related alignment sensitivities.  

Two separate Stark shift mitigation techniques were discussed.  Practical limitations of available power in a portable system  make the two-color, two-photon technique less appealing.  However, introduction of a Stark shift canceling laser could help reduce overall ac-Stark effects and allow for system fractional frequency instabilities as low $1\times10^{-15}$.  

The calculated BBR shift differs from extrapolated values.  The calculation results in two interesting temperatures: the magic tune out temperature, where the BBR polarizability is zero, and the temperature where the sensitivity to temperature variations is zero, effectively removing the BBR driven clock instability.   

\section*{acknowledgments} 
We thank Jordan Armstrong and Space Dynamics Laboratory for assistance in constructing the experimental apparatus.  Research preformed by M.S.S. was performed under the sponsorship of the US Department of Commerce, National Institute of Standards and Technology.
\appendix*

\begin{widetext}

\section{Cat's Eye Alignment Calculation}

Calculation of the intensity weighted average for the cat's eye case requires the Gaussian beam profile of the retro-reflected beam.   After retro-reflecting off of the cat's eye the beam re-enters the vapour cell with the following M matrix parameters:

\begin{equation}
\begin{pmatrix} X_f \\ Y_f \\ 1 \end{pmatrix} = \frac{1}{1+\Delta f}\begin{pmatrix}
    A&B&E \\
    C&D&F \\
    0&0&1+\Delta f
     \end{pmatrix} \begin{pmatrix} f\Delta X_i \\ \theta \\ 1 \end{pmatrix} ,
\end{equation}
where,
\begin{equation}
\begin{aligned}
A =& -(1+\Delta f-2\Delta d(N_2+\Delta d-N_2\Delta d+(N_2-1)\Delta f))), \\
B =& f(2(\Delta d-N_2(\Delta d-1))((N_2-1)\Delta d-1) +(2\Delta d(N_2+\Delta d-N_2 \Delta d)-1)N_1\\
&+(2(N_2-1)((N_2-1)\Delta d-1)+(2(N_2-1)\Delta d-1)N_2)\Delta f),\\
C =& -(-1+(N_2-1)\Delta d)f((2\Delta d-3\Delta f-1)\xi-\beta(1+\Delta f)),\\
D =&\frac{2\Delta d(1-\Delta d+\Delta f)}{f},\\
E =& -(1+\Delta f-2\Delta d(N_2+\Delta d-N_2\Delta d+N_1-\Delta dN_1+\Delta f(N_1+N_2-1))),\\
F =& \Delta d(\xi-2\Delta d\xi+3\Delta f\xi+\beta+\Delta f\beta).
\end{aligned}
\end{equation}
Here we utilize the same approximations as the flat mirror case, namely that all distances are much shorter than the incident beam Rayleigh length.  Even leveraging this approximation yields a complicated Gaussian function.  In order to simplify the analytical solution a few special cases were examined.

\begin{itemize}
\item[1. ] Setting $\Delta d=0$ yields
\begin{equation}
\overline{I}=\frac{\overline{I_{tot}}}{\overline{I_{tot}}(0)} =
 e^{\displaystyle\frac{-f^2(2x_i(1+\Delta f)+\xi+3\Delta f\xi+\beta+\Delta f\beta+(N_1+N_2)\theta-\Delta f\theta+(N_1+N_2)\Delta f\theta)^2}{3(1+\Delta f)^2w_0^2}}.
\end{equation}
This expression is shown in Figure \ref{fig:d0alignment} and is utilized to calculate Equations \ref{eq:cats_alignment_1} and \ref{eq:cats_alignment_2}.  

\item[2. ]  A numerical integral was performed using a quadrature method.  The numerical result is displayed in Figure \ref{fig:d0alignment} (b). 

\item[3. ]  The final case studied simplified the geometry to reduce the number of free parameters.  $\Delta f$ in known to have small impacts on the retro-reflected beam profile 
\cite{Snyder1975}.  Changes in $\Delta X_i$ and $\theta$ have similar effects on the beam propagation, a similar relationship exits between $\beta$ and $\Delta d$.  With this in mind another analytical case was examined where $\Delta X_i=0$, $\Delta f=0$ and $\beta=0$.  It was also necessary to expand the integrand in a Taylor series and ignore the higher order terms before final integration, yielding an average intensity of
\begin{equation}\label{eq:simple}
\begin{aligned} 
\overline{I} \approx& 1-\frac{f^2\xi^2+2f^2\xi\theta\left(N_1+N_2\right)+\theta^2f^2\left(N_2^2+2N_2N_1+N_1^2\right)}{3\omega_0^2} + \Delta d\left(4N_2+\frac{a}{f}+2N_1\right)+\\ &\Delta d^2\left(4-4N_2+\frac{88N_2^2}{9}+\frac{37a^2}{27f^2}-\frac{a}{f}+\frac{40aN_2}{9f}-2N_1+\frac{32N_2N_1}{3}+\frac{4aN_1}{3f}+4N_1^2\-\frac{2\pi^2\omega_0^4}{f^2\lambda^2}\right).
\end{aligned}
\end{equation}   

Taking the derivative of the above Equation w.r.t. either $\theta$ or $\xi$ yields the same result as the $\Delta d=0$ case. Figure \ref{fig:d0alignment} (b) shows average intensity given above plotted with the geometry presented in  Reference 
\cite{Martin2018} along with the numeric results for the same geometry.    

\end{itemize}
\end{widetext}

\newpage
\bibliography{clock_papers_3}

\end{document}